\documentclass[aps,pra,amsmath,amsfonts,superscriptaddress,reprint]{revtex4-2}

\usepackage{graphicx} 
\usepackage{braket} 
\usepackage{hyperref} 
\hypersetup{colorlinks=true, citecolor=blue, urlcolor=blue, linkcolor=blue}

\begin{document}


\title{Topological and nonlinearity-induced thermalization in a $\mathcal{PT}$-symmetric split-Langevin bath}
\date{\today}
\author{Andrew K. Harter}
\email{harter@lanl.gov}
\affiliation{Materials Physics \& Applications - Quantum Group, Los Alamos National Laboratory, Los Alamos, New Mexico 87545, USA}
\affiliation{Department of Physics, Indiana University Indianapolis, Indianapolis, IN 46202, USA}
\author{Donald J. Priour, Jr.}
\affiliation{Department of Physics \& Astronomy, Youngstown State University, Youngstown, OH 44555, USA}
\author{Daniel Sweeney}
\affiliation{Department of Physics, Indiana University Indianapolis, Indianapolis, IN 46202, USA}
\author{Avadh Saxena}
\affiliation{Theoretical Division and Center for Nonlinear Studies, Los Alamos National Laboratory, Los Alamos, New Mexico 87545, USA}
\author{Yogesh N. Joglekar}
\affiliation{Department of Physics, Indiana University Indianapolis, Indianapolis, IN 46202, USA}


\begin{abstract}
Open classical systems with balanced, separated gain and loss, called $\mathcal{PT}$  symmetric systems, have been extensively studied over the past decade. Here, we investigate the properties of a uniform, harmonic chain with spatially separated viscous loss and stochastic gain that are only statistically balanced. We show that such a “split Langevin” bath leads to either the absence of thermalization or non-equilibrium steady states with inhomogeneous temperature profile, both of which are understood in terms of normal modes of the chain. With a Su-Schrieffer-Heeger (SSH) chain, a canonical model with topological edge modes, we show that the steady-state properties reflect the topological phase of the underlying chain. We also show that nonlinearity stabilizes the amplifying modes in a harmonic chain, thereby leading to thermalization irrespective of the gain and loss locations. Our results expand the pool of possible realizations of non-Hermitian models to the stochastic domain.  
\end{abstract}
\maketitle


\section{Introduction}
\label{sec:intro}
More than twenty-five years ago, Bender and colleagues surprised the physics community by introducing a wide class of non-Hermitian Hamiltonians on an infinite line with purely real spectra~\cite{Bender:1998}. These Hamiltonians shared the key property that they were invariant under combined operations of parity ($\mathcal{P}$) and time-reversal ($\mathcal{T}$), i.e. they were $\mathcal{PT}$ symmetric. Intense mathematical research in the following decade clarified that $\mathcal{PT}$ symmetric Hamiltonians are a special case of pseudo-Hermitian~\cite{Mostafazadeh:2002,Mostafazadeh:2002a,Znojil:2008} Hamiltonians, and that an antilinear operator commuting with a Hamiltonian ensures a spectrum that is either purely real or has complex conjugate pairs~\cite{Mostafazadeh:2002b,Bender:2002}. It showed that when the spectrum is real, one can define a consistent quantum theory~\cite{Bender:2002a, *Bender:2004}, albeit one with a Hamiltonian-dependent, positive definite ($\mathcal{CPT}$) inner product~\cite{Mostafazadeh:2010} that also redefined the valid observables (self-adjoint operators) in the system; in particular, traditional, Dirac-Hermitian operators such as position, momentum, or spin {\it did not remain "observables"} under the $\mathcal{CPT}$ inner product.

Over the past decade, it has become clear that a non-Hermitian, $\mathcal{PT}$-symmetric Hamiltonian accurately describes an open system with balanced, separated gain and loss~\cite{Ruschhaupt:2005,El-Ganainy:2007,Makris:2008,Klaiman:2008,Joglekar:2013}. Starting with coupled optical waveguides~\cite{Guo:2009,Ruter:2010} and fiber loops~\cite{Regensburger:2012}, non-Hermitian $\mathcal{PT}$-symmetric systems have been experimentally realized with coupled optical rings~\cite{Hodaei:2014,Peng:2014}, electrical circuits~\cite{Schindler:2011,Leon-Montiel:2018,Wang:2020}, acoustics~\cite{Zhu:2014}, synthetic dimensions~\cite{Zhang:2020,Zhang:2020a}, as well as ultracold atoms~\cite{Li:2019}, superconducting circuit~\cite{Naghiloo:2019}, single NV center in diamond~\cite{Wu:2019}, entangled photons~\cite{Klauck:2019}, and a single trapped ion~\cite{Ding:2021,Quinn:2023}. These classical and quantum realizations have shown that judiciously engineering loss, (effective) gain, and the communication between them, leads to functionalities that are absent in closed systems~\cite{Miri:2019,Ozdemir:2019}. Thus, while compensating a loss locally with gain creates a closed, Hermitian system, compensating a loss with separated, balanced gain leads to non-Hermitian dynamics with surprising consequences.

Loss and gain also naturally occur in classical statistical systems. When the system - say, a chain of coupled oscillators – is in contact with a thermal reservoir, its resultant thermalization can be modeled by a Langevin thermostat~\cite{Langevin:1908,Lemons:1997}, i.e. a viscous drag force and accompanying local, random, thermal noise. The strength of the random force is determined by the damping coefficient and the bath temperature, and it leads to an equilibrium state where the loss and random gain are balanced on average~\cite{Kubo:1966,Lax:1960}. When multiple Langevin thermostats are present, indicating a system coupled to multiple reservoirs at different points, non-equilibrium steady states with thermal gradients across the system are observed~\cite{Rieder:1967,Lepri:2003}. In all of these examples, the losses are statistically and locally compensated by the random gain. We also note that $\mathcal{PT}$-symmetric harmonic chains connected to baths have also been studied~\cite{Zheng:2011}. In these active chains, however, include frictional losses that are is not accompanied by the noise, and a velocity-dependent gain, without requisite fluctuation effects. 

Here we consider Langevin configurations which are consistent with fluctuation dissipation theorem, and ask the following question: what is the fate of the system if the viscous loss and a statistically balanced random gain are spatially separated? To answer this question, we introduce a split-Langevin bath, and investigate the equilibration process in its presence. The plan of the paper is as follows. In Sec~\ref{sec:split}, we present numerical results for steady-state properties of the uniform chain with a split Langevin, that can be understood with the aid of stochastic and normal-mode analysis. In Sec.~\ref{sec:top} we show that this equilbriation is sensitive to the topological phase of the underlying Su-Schrieffer-Heegar (SSH) chain, with the topological edge modes hindering the process. In Sec.~\ref{sec:anh}, we discuss the influence of quartic and Hertz nonlinearities on the thermalization process in a uniform chain. The paper is concluded with a brief discussion in Sec.~\ref{sec:disc}.


\section{The Split Langevin Bath}
\label{sec:split}
Consider a one-dimensional, classical chain of $N+2$ identical particles with mass $M$ coupled by $N+1$ intermediate springs with spring constant $k$, where the auxiliary end masses (labeled $0$ and $N+1$) are fixed. The dynamics of such a system, in terms of the displacement from equilibrium $q_m(t)$ and the velocity $\dot{q}_m\equiv dq_m/dt$, are described by $N$ coupled differential equations. When this chain is immersed in a bath at temperature $T$, the resulting frictional force on each mass is given by $-M\gamma\dot{q}$ where $\gamma>0$ is the damping rate. The fluctuation-dissipation theorem~\cite{Kubo:1966} requires that each mass is also subject to a zero-mean, white-noise force $f(t)$ from the bath which satisfies $\braket{\braket{f(t)f(t')}}=2\gamma k_BT_0\delta(t-t')$ where $k_B$ is the Boltzmann constant. $T_0$ is the temperature of the bath, and $\braket{\braket{\cdot}}$ denotes statistical averaging. Due to this random gain, the time-averaged energy density of a long chain reaches a steady state value equal to $k_B T$ that is consistent with the equipartition theorem. Note that the equilibration occurs, on a much longer time scale, if the Langevin thermostat is coupled only to the end masses.

\begin{figure}[h!]
\centering
\includegraphics[width=0.9\columnwidth]{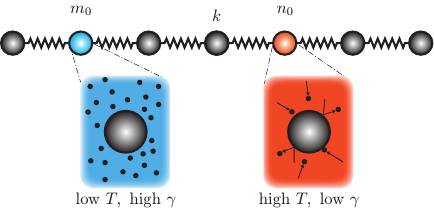}
\caption{\label{fig:split-bath} Schematic rendering of a split-Langevin bath for a harmonic, uniform chain with $N=7$ sites. The viscous force (loss) is only at site $m_0$ (shown in blue) and the random fluctuations (gain) are only at site $n_0$ (shown in red). Using a vanishingly low-temperature bath satisfies the fluctuation-dissipation theorem constraint at the loss site $m_0$, as does a vanishingly small $\gamma$ at the random-gain site $n_0$.}
\vspace{-3mm}
\end{figure}

Motivated by the idea of spatially separating loss and gain locations, we consider the split Langevin configuration, see Fig.~\ref{fig:split-bath}. The drag force $-M\gamma\dot{q}$ acts only on a mass at site $m_0$ (shown in blue), with vanishingly small random force. Conversely, a random force $f(t)$ with nonzero variance acts on the mass at site $n_0$ (shown in red), but there is little-to-no drag at that location. All other masses (shown in black) are isolated from the environment. Physically, such an arrangement is made possible by a vanishingly small temperature at site $m_0$ which minimizes the random force at that site, and a vanishingly small damping at site $n_0$. What are the dynamics of such a chain? Will it always reach a steady state? How do its steady-state properties depend on $m_0$ and $n_0$?

The chain dynamics are described by the following set of $N$ coupled differential equations,
\begin{equation}
\label{eq:splitm}
\ddot{q}_m=\omega^2(q_{m-1}-2q_m+q_{m+1})-\gamma\dot{q}_m\delta_{mm_0}+f(t)\delta_{mn_0}, 
\end{equation}
which make explicit the separation of the viscosity-dominated effects and the temperature-dominated effects. Open boundary conditions are imposed by constraints $q_0 (t)=0=q_{N +1}(t)$. Without loss of generality, we use units where $M=1$ sets the mass scale, $\sqrt{k/M}=\omega=1$ sets the inverse-time scale, and $k_BT_0=1$ sets the energy scale or equivalently, $T_0=1$ sets the temperature scale; they generate a thermal length-scale $q_T\equiv\sqrt{k_BT_0/M\omega^2}=1$. We solve Eq.(\ref{eq:splitm}) via velocity-Verlet integration~\cite{Verlet:1967,Swope:1982}, and obtain the time-averaged kinetic energy for each mass
\begin{equation}
\label{eq:ke}
\mathcal{E}_m\equiv\lim_{\tau\rightarrow\infty}\frac{1}{\tau}\int_0^\tau dt'\frac{1}{2}M\dot{q}_m^2(t').
\end{equation}
In the following, we use the site-dependent kinetic energy $\mathcal{E}_m$ or the local temperature $T_m\equiv2\mathcal{E}_m/k_B$ interchangeably. For the usual Langevin model ($m_0=n_0$), the chain equilibrates to a uniform temperature profile, i.e. $\mathcal{E}_m/(k_BT_0)\rightarrow 1/2$ for all $m$. When $m_0\neq n_0$, the split-Langevin system either reaches a steady state with a site-dependent kinetic energy, or it fails to reach a steady state and the time-averaged kinetic energy diverges with the integration time $\tau$. If the chain reaches a steady state, we define the dimensionless mean-temperature of the chain as
\begin{equation}
\label{eq:meanT}
\braket{T}=\frac{2}{Nk_BT_0}\sum_{m=1}^N\mathcal{E}_m\equiv\frac{1}{N}\sum_{m=1}^N T_m.
\end{equation}

\begin{figure}
\centering
\includegraphics[width=\columnwidth]{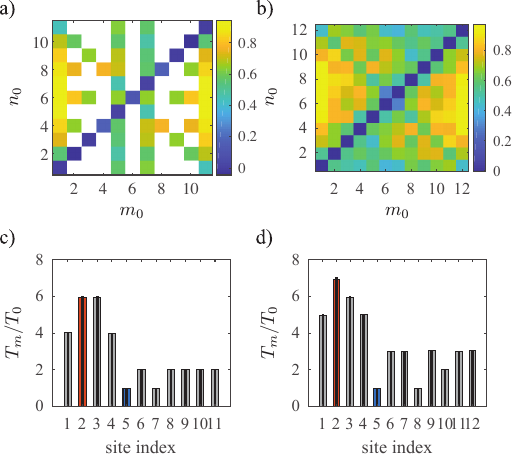}
\caption{\label{fig:phasetm} 
Mean-temperature profile for a split-Langevin chain with $N$ masses, drag at site $m_0$ and random gain at site $n_0$. All temperatures are measured in units with $k_BT_0=1$. (a) $\log_{10}\braket{T}$ for an $N=11$ chain shows that mean temperatures vary over a factor of ten; white squares denote configurations where the chain does not reach a steady state. (b) for an $N=12$ chain, steady state is reached for all configurations. In (a) and (b), the largest mean temperature $\braket{T}$ is reached when the viscous drag is at end points and the random gain is in the center. Local temperature profiles for an $N=11$ (c) or $N=12$ (d) chain with loss at $m_0=5$ (blue bar) and random gain at $n_0=2$ (red bar). The gray boxes indicate the theoretical results, whereas the black bars indicate the values resulting from direct numerical integration and averaging.}
\vspace{-5mm}
\end{figure}

Figure~\ref{fig:phasetm} shows typical results for $\log_{10}\braket{T}$ obtained from numerical simulation of Eq.(\ref{eq:splitm}) as a function of the loss site $m_0$ and the random-gain site $n_0$ for chains with $N=11$ (a) and $N=12$ (b) sites respectively. The white squares denote configurations where the chain does not reach a steady state. The salient features of these results are as follows. When $ m_0=n_0$, we obtain the usual Langevin result, namely the chain equilibrates to a uniform, constant temperature. When $m_0\neq n_0$, the chain {\it does not reach a steady state if and only if $m_0$ and $n_0$ are co-prime and $m_0$ is a nontrivial factor of $N+1$}. Thus, when $N=11$ and the loss is on site $m_0 =3$, the chain reaches a steady state if and only if the random gain is on site $n_0=\{3,6,9\}$. On the other hand, for an $N=12$ chain, all loss-gain configurations lead to steady-state results because $N + 1$ is a prime number. We also note that the highest steady-state temperature -- a ten-fold increase from the traditional Langevin result – is achieved when the loss is on the end masses and the gain is at the center of the chain. The results are symmetric under reflection, i.e. $\braket{T}(m_0,n_0)=\braket{T}(N+1-m_0,N +1-n_0)$, but not under the exchange of loss and random-gain locations, i.e. $\braket{T}(n_0,m_0)\neq\braket{T}(m_0,n_0)$. 

Figures~\ref{fig:phasetm}c and~\ref{fig:phasetm}d show the steady-state local temperature for a configuration with random gain on site $n_0=2$ (red bar) and the viscous drag on site $m_0=5$ (blue bar) for chains with $ N=11$ sites (c) and $N=12$ (d) sites. The black bars are results obtained from numerical integration of Eq.(\ref{eq:splitm}) with cutoff $\tau=(2\pi/\omega)\times 10^5$ and time-step $\Delta t=0.1(2\pi/\omega)$; they remain unchanged when the cutoff is increased or the time-step is reduced. The gray boxes denote analytical results that are obtained from the steady-state momentum covariance matrix (Appendix~\ref{sec:stoc}). For both chains, the non-equilibrium steady states have heterogeneous temperature profiles that are dictated by the loss and random-gain locations. In particular, we see that although the random-gain site is the hottest, and the viscous drag site is the coolest, the site temperatures at other locations are not, in general, monotonic in the distance from the gain or the loss location. These results are valid for any chain size $N$, and imply that a split Langevin bath offers ways to engineering non-equilibrium steady states with complex temperature profiles.

To understand the long-term properties of the chain, we write Eq.(\ref{eq:splitm}) as a stochastic differential equation~\cite{Gardiner:2009}
\begin{equation}
\label{eq:sde}
d{\bf x} =-A{\bf x}dt+ B d{\bf W},
\end{equation}
where ${\bf x}=({\bf q},\dot{\bf q})^T$ is a column vector with positions and velocities. The $2N\times2N$ matrices $A$ and $B$ are given by
\begin{equation}
A = \begin{bmatrix} 0_N & -1_N \\ K/M & \gamma P_L \end{bmatrix}\,, \quad
B = \begin{bmatrix} 0_N & 0_N \\ 0_N & \sqrt{2\gamma k_BT_0} P_G \end{bmatrix} \,,
\end{equation}
where $K$ is the tridiagonal, spring constant matrix, $P_L$ is the projection operator onto the loss sites, and $P_G$ is the projection operator onto the random-gain region. The stochastic term $d{\bf W}=(0,\cdots,0,dw_1,\cdots,dw_N)^T$ has $N$ independent, identically distributed variables $dw_i$ with zero mean and unit variance. 

The covariance matrix $\sigma(t)\equiv\braket{\braket{{\bf x}(t){\bf x}^T(t)}}$ satisfies the  equation~\cite{Gardiner:2009} $\dot{\sigma} = -(A\sigma+\sigma A^T)+BB^T$ with solution
\begin{equation}
\label{eq:sigma}
\sigma(t)=\int_0^t e^{-A(t-t')} BB^T e^{-A^T(t-t')} dt'. 
\end{equation}
{\it If the system reaches a steady state}, then the steady-state solution $\sigma_{ss}$ satisfies the continuous Lyapunov equation~\cite{Bartels:1972}, $A\sigma_{ss} + \sigma_{ss} A^T = BB^T$. To obtain the steady-state criterion, we explicitly evaluate Eq.(\ref{eq:sigma}) by using spectral decomposition of the non-Hermitian, real-valued matrix $A=\sum_{n=1}^{2N}\lambda_n\ket{L_n}\bra{R_n}$, and find 
\begin{equation} 
\label{eq:sigmamn}
\sigma_{mn}=\langle L_m|B B^T|L_n\rangle\left[\frac{1 - e^{-(\lambda_m+\lambda_n^*)t}}{\lambda_m+ \lambda_n^*}\right].
\end{equation}
Here $\bra{L_n}A=\lambda_n\bra{L_n}$ denotes left (row) eigenvector of $A$ with eigenvalue $\lambda_n$, $A\ket{R_n}=\lambda_n\ket{R_n}$ is the corresponding right eigenvector, and $\ket{L_n}=\bra{L_n}^\dagger\neq \ket{R_n}$ is the Hermitian conjugate of the left eigenvector. Since $A$ is a real matrix, the eigenvalues $\lambda_n$ occur in complex conjugate pairs. It follows that Eq.(\ref{eq:sigmamn}) will reach a steady state if and only if all eigenvalues have positive real parts, $\Re\lambda_m>0$, or $\bra{L_m}BB^T\ket{L_n}=0$ when $\Re\lambda_m=0$. Using normal modes $\ket{\alpha}$ of the spring-constant matrix $K$ (Appendix~\ref{sec:stoc}), it follows that the system does not equilibrate when a mode decouples from the loss region, $P_L\ket{\alpha}=0$ while remaining coupled to the gain region, $P_G\ket{\alpha}=0$. For a single split-Langevin bath, this criterion reduces to the joint requirements that $m_0,n_0$ are co-prime and $m_0$ is a nontrivial factor of $N + 1$ (Fig.~\ref{fig:phasetm}a). In particular, such system always reaches steady state if $N+1$ is a prime (Fig.~\ref{fig:phasetm}b).

We have seen that the equilibration of the system is determined by the overlap of the isolated-system normal modes with the split-Langevin bath. In the next section, we show that topological properties of these normal modes significantly affect the approach to equilibration.


\begin{figure*}
\centering
\includegraphics{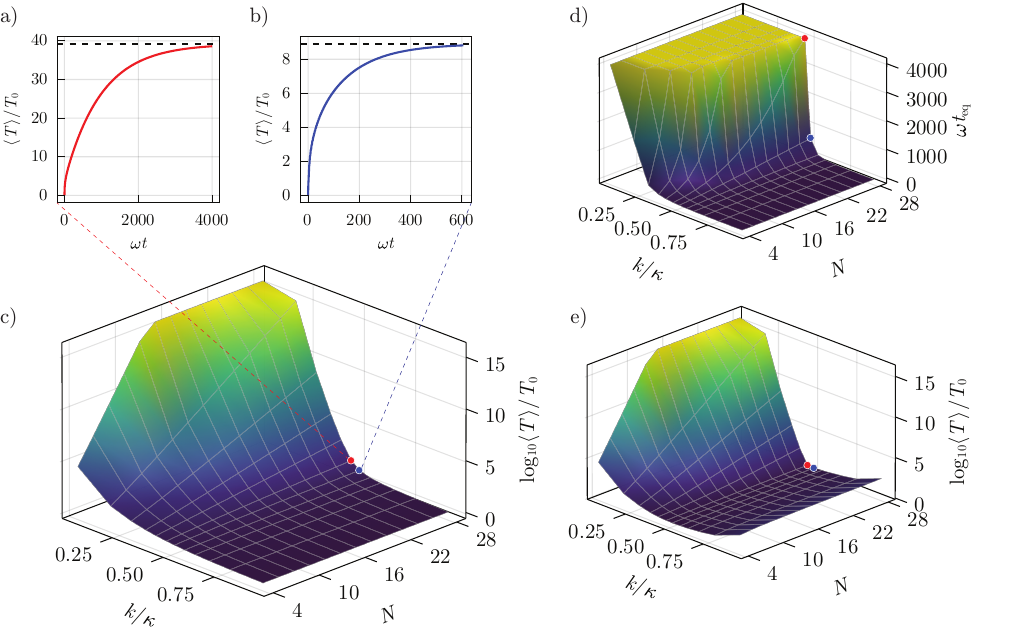}
\caption{\label{fig:ssh-tnk}Transient properties and equilibration of an even SSH chain with split-Langevin. (a) in the topological phase, $k/\kappa=0.48$, reaching equilibrium takes long time for the mean temperature $\braket{T}(t)$. (b) in the trivial phase, $k/\kappa=0.52$, the mean temperature reaches a much lower $\braket{T}$ in a tenth of the time. (c) $\log_{10}\braket{T}(k/\kappa,N)$ shows that the equilibrium temperature $\braket{T}$ is weakly sensitive to both parameters in the topologically trivial phase given by $k/\kappa>0.5$, while in the topological phase, $k/\kappa<0.5$, it increases by orders of magnitude and is highly sensitive to $N$. Thus, presence of topological states inhibits rapid thermalization. The red and blue points indicate system parameters for panels a and b. (d) the equilibration time $\omega t_{\mathrm{eq}}(k/\kappa,N)$ also increases rapidly when the system enters the topological phase at $k/\kappa\leq 0.5$ and acquires a clear $N$-dependence; region where $t_\mathrm{eq}$ exceeds the simulation cutoff $\tau_{\max}$ is shown by a yellow plateau. For (a)-(d), the SSH chain has a loss and random gain in each unit cell, i.e. a total of $N/2$ split-Langevin baths. (e) SSH chain with {\it single} split-Langevin bath at its ends shows qualitatively same behavior for $\log_{10}\braket{T}(k/\kappa,N)$; only the equilibrium temperatures have a slight rise at the edge near $k/\kappa = 1$.}
\end{figure*}

\section{Topological Effects}
\label{sec:top}

\begin{figure}
\centering
\includegraphics[width=\columnwidth]{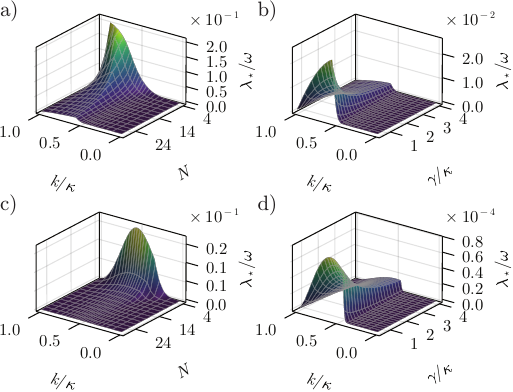}
\caption{\label{fig:minrea} Behavior of $\lambda_\star\equiv\min\Re\lambda_n$, obtained from matrix $A$, as a proxy for inverse thermalization time $1/t_\mathrm{eq}$. (a) for an SSH chain with split-Langevin in each unit cell, $\lambda_\star(k/\kappa,N)$ is highly suppressed for $k/\kappa<0.5$ indicating slow thermalization in the topological phase. In the trivial phase, $\lambda_\star$ decreases with increasing $N$ but on average remains higher than its value in the topological phase. (b) the same chain ($N=28$) shows thermalization slowdown with increasing $\gamma$ in the trivial phase, with little variation in the topological phase. (c)-(d) an SSH chain with a single split-Langevin at its ends shows qualitatively, somewhat the same behavior; however, in (c), increased values of $\lambda_\star$ for small systems persist into the region where $k/\kappa < 0.5$.}
\end{figure}

As a canonical, one-dimensional model with topological phase, we consider the Su-Schrieffer-Heeger (SSH) chain~\cite{Su:1979,Rudner:2009,Zeuner:2015}. It consists of identical masses coupled by springs with spring constants alternating between $k$ and $k'$, thereby forming a unit cell with two masses in it. Such open chain with integer number of unit cells has two, zero energy topological edge modes localized at both edges when $k<k'$, while for $k\geq k'$, all modes are extended. First, we consider an SSH chain with alternating loss and random-gain, so that both are present in every unit cell, and use the dimensionless ratio $k/\kappa\in [0,1]$ to distinguish the topological phase ($k/\kappa<0.5$) from the trivial phase ($k/\kappa\geq 0.5$), with $\kappa=k+k'$. We obtain the dimensionless, mean, time-dependent temperature 
\begin{equation}
    \label{eq:transientT}
    \langle T\rangle (t)=\frac{2}{Nk_BT_0}\sum_{m=1}^N \int_{0}^t dt' \frac{1}{2}M\dot{q}_m^2(t')
\end{equation}
at times $t\leq\tau_{\max}=4000/\omega$. We define the equilibration time $t_\mathrm{eq}\leq\tau_{\max}$ as the shortest time when mean temperature reaches within 1\% of its equilibrium value. Results of such analysis are summarized in Fig.~\ref{fig:ssh-tnk}. For a chain with $N=28$ masses and $k/\kappa=0.48<0.5$, the chain is in topological phase and has robust edge modes. Figure~\ref{fig:ssh-tnk}a shows that in this case, the mean temperature $\langle T\rangle(t)$ slowly rises to its equilibrium value (black dashed line). In contrast, when $k/\kappa=0.52>0.5$, the chain is in topologically trivial phase. Fig.~\ref{fig:ssh-tnk}b shows that the $\langle T\rangle(t)$ reaches its new, smaller equilibrium value (black dashed line) in a tenth of the time. Figure~\ref{fig:ssh-tnk}c shows the equilibrium temperature $\log_{10}\braket{T}$ as a function of $k/\kappa$ for even chains with sizes $6\leq N\leq 28$. We see that the topological phase has orders-of-magnitude higher $T_\mathrm{eq}$ values that increases uniformly with $N$, while in the topologically trivial phase, $k/\kappa>0.5$, the equilibrium temperature $\sim O(1)$ remains insensitive to the chain size. In Fig.~\ref{fig:ssh-tnk}d, we plot the time to equilibration $\omega t_{\mathrm{eq}}(k/\kappa,N)$. In the topologically trivial phase $k/\kappa>0.5$, this time is approximately independent of the chain size and relatively short ($\lesssim 400$). In contrast, when the chain has topological edge modes, $k/\kappa<0.5$, this time becomes strongly $N$ dependent, and rises rapidly. For long chains deep in the topological phase $k/\kappa\ll 0.5$, it exceeds the simulation cutoff $\omega\tau_{\max}=4000$; this is indicated by the (yellow) plateau. The red and blue points correspond to $k/\kappa=0.5\pm 0.02$ and $N=28$, and show that a change of $<10\%$ in spring constant across the topological phase boundary at $k/\kappa=0.5$ leads to tenfold increase in the equilibration time (Fig~\ref{fig:ssh-tnk}a and Fig.~\ref{fig:ssh-tnk}b). Since the $(N-2)$ bulk states of the SSH chain do not change significantly across the phase boundary, this increase is solely due to the emergence of two, robust, edge-localized modes that signify topologically nontrivial phase. 

This key result --- topological edge states inhibit thermalization in the split-Langevin model --- holds when, instead of having a loss and random gain in each unit cell, we consider an even SSH chain with loss at one end ($m_0=1$) and random gain at the other end ($n_0=N$) so that all normal modes interact with both gain and loss. The resulting equilibrium temperature dependence $\log_{10}\braket{T}(k/\kappa,N)$, Fig.~\ref{fig:ssh-tnk}e, shows qualitatively similar behavior across the topological phase boundary. It is important to note that the equilibrium temperatures here are much smaller, see Fig.~\ref{fig:ssh-tnk}c, due to the presence of a single loss and random gain. To understand the equilibration time, Fig.~\ref{fig:ssh-tnk}d, we examine the minimum real part of the eigenvalues of the matrix $A$, $\lambda_\star\equiv\min\Re\lambda_n$, as it dictates the slowest-decaying term in the time-dependent covariance matrix, Eq.(\ref{eq:sigmamn}). For an SSH chain with split-Langevin in each cell, Fig.~\ref{fig:minrea}a shows that $\lambda_\star(k/\kappa,N)$ is large and decreases with $N$ when the system is in the trivial phase, $k/\kappa<0.5$ whereas in the topological phase, $\lambda_\star$ is highly suppressed indicating divergent relaxation times. Figure~\ref{fig:minrea}b shows that as the strength $\gamma$ of the random gain and loss is increased, $\lambda_\star$ decreases in the topologically trivial phase. In contrast, in the presence of topological edge modes, the slow thermalization process, indicated be highly suppressed $\lambda_\star$ is largely unaffected by variations of $\gamma$. Figure~\ref{fig:minrea}c and Fig.~\ref{fig:minrea}d show corresponding results for an SSH chain with a single split-Langevin. Beyond an overall suppression of $\lambda_\star$ due to the presence of a single split-Langevin, the results are similar to those for an SSH chain with $N/2$ split-Langevins. 


\section{Effects of nonlinearity}
\label{sec:anh}
\begin{figure}[b]
\centering
\includegraphics[width=\columnwidth]{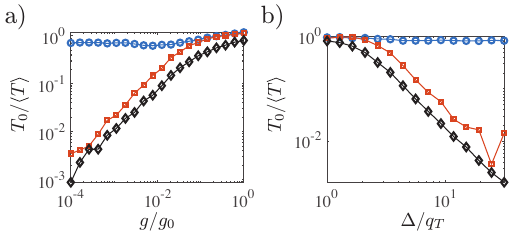}
\caption{\label{fig:tanharm} Nonlinearity-induced thermalization in an $N=9$ chain with random gain at site $n_0=1$ and loss at $m_0=3$ (blue circles), or $m_0=\{2,5\}$ (red squares, black diamonds). The harmonic chain does no thermalize in the latter cases. (a) Inverse mean-temperature $\braket{T}^{-1}$ as a function of the quartic-potential strength $g$ shows that when $g>0$, for $m_0=\{2,5\}$, chain reaches a steady-state with decreasing $\braket{T}$, reaching unity when $g/g_0\sim1$. (b) $\braket{T}^{-1}$ as a function of precompression $\Delta$ in the Hertzian contact force shows that the non-thermalizing cases, $m_0=\{2,5\}$, with divergent $\braket{T}$ at large $\Delta/q_T$ become stable when $\Delta$ is reduced due to normal mode mixing. For both nonlinearities, $\braket{T}$ is roughly constant if the chain reaches a steady-state in the harmonic limit, $m_0=3$.}
\end{figure}

Lastly, we discuss the effect of spring nonlinearity on the fate of thermalization. 
The potential energy of the  chain in Fig.~\ref{fig:split-bath} is given by the quadratic form $U_0({\bf q})={\bf q}^TK{\bf q}/2$. When spring nonlinearities are taken into account, the potential energy changes to
\begin{equation}
\label{eq:anh1}
U({\bf q})=U_0({\bf q})+g\sum_{m=1}^{N-1}(q_{m+1}-q_m)^4,
\end{equation}
where $g$ is measured in units of $g_0\equiv k_BT_0/q_T^4=1$. The quartic interaction mixes the normal modes of the harmonic chain, and therefore the system reaches a steady state for all configurations of the split-Langevin bath. To demonstrate this phenomenon, we obtain the dimensionless, steady-state, mean temperature $\braket{T}$ by numerically integrating the equations of motion with the nonlinear force in the presence of a split Langevin. Figure~\ref{fig:tanharm}a shows $\braket{T}^{-1}(g)$ for an $N=9$ uniform chain with gain at $n_0=1$ and three different locations for the viscous loss, $m_0=\{2,3,5\}$. If the chain reaches a steady-state in the limit $g=0$, introducing the nonlinearity does not significantly change its temperature $\braket{T}$ ($m_0=3$, blue circles). In contrast, when the chain does not thermalize in the harmonic limit, $\braket{T}^{-1}(0)=0$, we see that increasing $g$ leads to thermalization, with a large nonlinearity $g/g_0\sim 1$ required to get $\braket{T}\sim T_0$ ($m_0=2$, red squares; $m_0=5$, black diamonds).  

Lastly, we consider a model with Hertzian potential. The contact force between two precompressed elastic spheres with locations $q_1$ and $q_2$ can be written as~\cite{Landau:1986, Porter:2015}
\begin{equation}
\label{eq:anh2}
F_{\mathrm{Hz}}(q_1,q_2)=Mg_{\mathrm{Hz}}\Theta(Q_{12})Q_{12}^{3/2},
\end{equation}
where $Q_{12}\equiv\Delta-|q_1-q_2|$ is change from precompressed center-to-center distance $\Delta$, $M$ is mass of each sphere, and $g_{\mathrm{Hz}}$ is the strength of the contact force. The Heaviside function $\Theta$ represents the fact that when the center-of-mass distance is larger than its precompressed value $\Delta$, the spheres do not exert any force on each other; on the other hand, as the spheres get closer, they exert a repulsive force that is stronger than the Hooke’s law. For very small displacements $|q_1-q_2|\ll\Delta$, the Hertzian force reduces to the harmonic model with a spring constant $k_{\mathrm{Hz}}=3Mg_{\mathrm{Hz}}\sqrt{\Delta}/2$. Figure~\ref{fig:tanharm}b shows the inverse mean temperature of a chain of $N=9$ such spheres in the presence of the same split-Langevin as in Fig.~\ref{fig:tanharm}a. Once again, we see that if the chain thermalizes in the harmonic limit, $\Delta/q_T\gg 1$, then introducing nonlinearity by decreasing $\Delta$ does not affect the equilibrium temperature value ($m_0=3$, blue circles). However, if the harmonic chain does not thermalize, the Hertzian nonlinearity stabilizes the chain, with the steady-state temperature $\braket{T}\sim T_0$ when $\Delta/q_T\sim 1$ ($m_0=2$, red squares; $m_0=5$, black diamonds).


\section{Discussion}
\label{sec:disc}

Recent studies of systems with balanced gain and loss, or $\mathcal{PT}$ symmetric systems, have shown that spatially separating the gain from loss leads to new functionalities that are absent when the gain locally compensates for the loss. Motivated by this physical picture, we have introduced the concept of a split-Langevin bath, where the viscous loss in one location is compensated by a spatially separated random gain that is statistically balanced. 

We have shown that steady states with non-trivial local temperature profiles occur in such configurations, whereas a regular Langevin bath leads to a uniform temperature. In a complementary picture, the split-Langevin bath permits selective heating of some normal modes. We have shown that the mean steady-state temperature and the time to reach it are both qualitatively affected across a topological transition, where the emergence of robust, topological edge modes hinders the efficient redistribution of energy in the lattice and thereby leads to higher thermalization temperatures. We have also shown that nonlinearity, in general, stabilizes split-Langevin configurations with no-steady-state in the harmonic limit. 

Granular crystals consisting of macroscopic, elastic spheres have emerged as a promising candidate for studying single-particle and nonlinear dynamics in a controlled and tunable setting~\cite{Porter:2015,Chong:2017}. One dimensional granular chains have been used to observe band structure~\cite{Porter:2015}, solitary waves~\cite{English:2005,Sen:2008}, interface shock waves~\cite{Molinari:2009}, breathers~\cite{Boechler:2010,Chong:2014}, and acoustic switches~\cite{Li:2014}. The ability to address, excite, and track individual particles in such chains suggests that a split-Langevin bath may be engineered in such chains more easily than in nonlinear spring and mass chains.


\section*{acknowledgements}
This work was supported by Office of Naval Research Grant No. N00014-21-1-2630 (YJ) and the U.S. Department of Energy (AKH, AS).


\appendix

\section{Thermalization Analysis in the Normal-mode Picture}
\label{sec:stoc}

In this section, we derive the covariance matrix and criteria for thermalization for a harmonic chain by using its normal modes. Let us denote the covariance matrix in block diagonal form~\cite{Rieder:1967}
\begin{equation}
\label{eq:sigmablock}
\sigma\equiv\braket{\braket{{\bf x}{\bf x}^T}}=\begin{bmatrix}
\braket{\braket{{\bf q}{\bf q}^T}} & \braket{\braket{{\bf q}\dot{\bf q}^T}}\\
\braket{\braket{\dot{\bf q}{\bf q}^T}} & \braket{\braket{\dot{\bf q}\dot{\bf q}^T}}
\end{bmatrix} =
\begin{bmatrix}
\sigma_q & c\\
-c & \sigma_{\dot{q}}
\end{bmatrix}.
\end{equation}
The trace of the velocity correlation matrix $\sigma_{\dot{q}}$ can be used to obtain the mean temperature $\braket{T}$. We can integrate Eq.(\ref{eq:sigma}) up to a time $t_\mathrm{eq}$, but in cases where the system does not thermalize, the matrix $A$ becomes badly conditioned so that its inversion, required to solve the steady-state equation $A\sigma_{ss}+\sigma_{ss}A^T=BB^T$, becomes numerically unstable. In terms of its position, velocity, and cross-correlation blocks, the steady-state constraint becomes 
\begin{gather}
\label{eq:steadyse}
(c K - K c)/M + \gamma (P_L \sigma_{\dot{q}} + \sigma_{\dot{q}} P_L) =2\gamma k_BT_0 P_G, \\
\label{eq:steadysw}
K\sigma_q/M -\gamma P_L c-\sigma_{\dot{q}} = 0. 
\end{gather}
Let $\ket{\alpha}$ denote the normal modes of the matrix $(K/M)\ket{\alpha}=\omega_\alpha^2\ket{\alpha}$ where $\omega_\alpha=2\omega\sin\left[\pi\alpha/2(N+1)\right]$ are the normal mode eigenvalues for a uniform chain and $\braket{m|\alpha}=\sqrt{2/(N+1)}\sin\left[\pi\alpha m/(N+1)\right]$ are corresponding eigenfunctions consistent with open boundary conditions, and $1\leq\alpha\leq N$~\cite{Ashcroft:2022}. Taking the trace of Eq.(\ref{eq:steadyse}) in the normal-mode basis gives
\begin{equation}
\label{eq:ss}
\braket{\alpha|P_L|\alpha} k_BT_\alpha+\sum_{\beta\neq\alpha}\braket{\alpha|P_L|\beta}\braket{\beta|\sigma_{\dot{q}}|\alpha}=k_BT_0\braket{\alpha|P_b|\alpha},
\end{equation}
because diagonal entries of $\sigma_{\dot{q}}$ give the eigenmode temperature. It is clear that Eq.(\ref{eq:ss}), a necessary condition for a steady-state result, cannot be satisfied if $P_L\ket{\alpha}=0$ and $P_G|\alpha\rangle\neq 0$.

When all normal modes couple with the loss region, $P_L\ket{\alpha}\neq0$, the steady-state temperature of a normal mode is given by
\begin{equation}
\label{eq:talpha}
k_BT_\alpha=k_B T_0\frac{\braket{\alpha|P_G|\alpha}}{\braket{\alpha|P_L|\alpha}}-\sum_{\beta\neq\alpha} \frac{\braket{\alpha|P_L|\beta}}{\braket{\alpha|P_L|\alpha}}\braket{\beta|\sigma_{\dot{q}}|\alpha}.
\end{equation}
For the standard Langevin bath this simplifies to $T_\alpha=T_0$, i.e. all normal modes have the same temperature. For a split-Langevin bath, this result enhanced by ratio of normal-mode weights on the random-gain site and the viscous-loss site. Additionally, the second term in Eq.(\ref{eq:talpha}) also contributes off-diagonal velocity correlations in the matrix $\sigma_{\dot{q}}$. 

\begin{figure}
\vspace{5mm}
\centering
\includegraphics[width=\columnwidth]{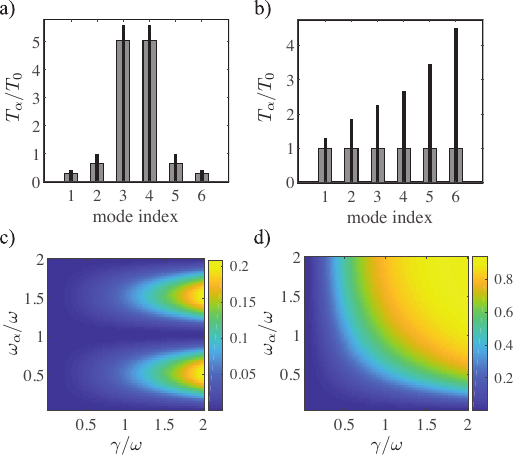}
\caption{\label{fig:tkcomp} Normal-mode temperatures $k_BT_\alpha$ for a chain with $N=6$ obtained from Eq.(\ref{eq:talpha}). (a) for $m_0=1, n_0=2$, the off-diagonal contribution to $T_\alpha$ (black bars) is small and the temperature profile is symmetric in the mode index (gray boxes). (b) for $m_0=1, n_0=N$, the off-diagonal contribution increases with the mode index (black bars: exact result; gray boxes: diagonal contribution). Normalized off-diagonal contribution $\Delta T(\omega_\alpha,\gamma)$ for a chain with $N=60$. (c) when $m_0=1, n_0=2$, $\Delta  T$ is small and symmetric about the frequency center. (d) for $m_0=1, n_0=N$ , $\Delta T$ increases monotonically with the normal-mode frequency. Temperature is in units of $k_BT_0=1$ and $\omega_\alpha,\gamma$ are in units of $\omega=1$.}
\end{figure}

Figure~\ref{fig:tkcomp}a shows the normal-mode temperatures $k_BT_\alpha$ for an $N=6$ site chain with loss on the first site, $m_0=1$, and random gain on the second site, $n_0=2$. The gray boxes denote the first term in Eq.(\ref{eq:talpha}) and the black bars are the exact result. In this case, contribution from off-diagonal terms $\braket{\beta|\sigma_{\dot{q}}|\alpha}$ is negligible. In contrast, for $m_0=1$ and $n_0=N$, Fig.~\ref{fig:tkcomp}b shows that the $T_\alpha$-profile has a strong contribution from other modes that monotonically increases with the mode index $\alpha$. We plot the normalized off-diagonal contribution 
\begin{equation}
    \label{eq:delta}
    \Delta T(\omega_\alpha,\gamma)\equiv \frac{1}{k_B T_{\alpha}}\sum_{\beta\neq\alpha}\frac{\braket{\alpha|P_L|\beta}}{\braket{\alpha|P_L|\alpha}}\braket{\beta|\sigma_{\dot{q}}|\alpha},
\end{equation}
for an $N= 60$ site chain with adjacent loss and random-gain sites, Fig.~\ref{fig:tkcomp}c, and farthest-removed loss and random-gain sites, Fig.~\ref{fig:tkcomp}d. As is expected, $\Delta T$ increases with the random-gain strength. For adjacent loss and gain, it is symmetric about the frequency-range center, and overall small in magnitude, as seen in Fig.~\ref{fig:tkcomp}a. On the other hand, when the loss and the random gain are maximally separated, the normalized difference $\Delta T$ increases with the mode frequency and is, on average, larger in magnitude, as seen in Fig.~\ref{fig:tkcomp}b. In a standard Langevin bath, the thermal energy is equally distributed among all modes. Figure~\ref{fig:tkcomp} shows that by judiciously engineering the split-Langevin bath, one can selectively heat up phonons in the band middle (a,c) or at the band edge. 

Next, we obtain constraints on the eigenvalues $\lambda_n$ and left (right) eigenvectors $\bra{L_n} (\ket{R_n})$ of matrix $A$. When the eigenvalue problem $A\ket{R_n}=\lambda_n\ket{R_n}$ is cast in terms its $N\times N$ blocks, the $2N$ dimensional eigenvectors $\ket{R_n}$ can be written as $\ket{R_n}=(\ket{v_n},-\lambda_n\ket{v_n})^T$ where the $N$-dimensional (column) vector $\ket{v_n}$ satisfies the equation 
\begin{equation}
\label{eq:evalue}
\left(K/M -\lambda\gamma P_L + \lambda^21_N\right)\ket{v_n} = 0 \,.
\end{equation}
It follows that the $2N$ eigenvalues of the matrix $A$ are given by the $N$ complex-conjugate pairs $(\lambda,\lambda^*)$ resulting from Eq.(\ref{eq:evalue}). In the absence of drag and thermal noise, the solutions of Eq.(\ref{eq:evalue}) are $\lambda=\pm i\omega_n$ where $\omega_n$ are the normal mode frequencies of matrix $K/M$. In this case, all eigenvalues are imaginary and no steady-state can be reached. A leading-order correction to the eigenvalues as $\gamma$ is increased from zero implies that
\begin{equation}
\label{eq:evalue2}
\lambda_n=\pm i\omega_n+\frac{\gamma}{2}\braket{v_n^0|P_L|v_n^0} \,.
\end{equation}
where $\ket{v^0_n}$ are the phonon eigenmodes with eigenvalue $\omega_n$. Thus, a steady state is achieved if all normal modes couple to the loss region, $\braket{v_n^0|P_L|v_n^0}\neq 0$. If a particular mode {\it does not project onto the loss region}, the steady-state constraint from Eq.(\ref{eq:sigmamn}) requires $\braket{L_n|BB^T|L_m}=0$. By writing $\bra{L_n}|=(\bra{v_n}K,\lambda_n\bra{v_n})$, the steady-state constraint reduces to $\braket{v_n|P_G|v_m}=0$, i.e. the mode must decouple from the random-gain region as well for the system to reach a (time-averaged) steady state. 

We note in the passing that the site-dependent kinetic temperature $T_m$ is also obtained from the steady-state covariance matrix,
\begin{equation}
\label{eq:tm}
k_B T_m=\sum_{\alpha,\beta=1}^N\braket{m|\alpha}\braket{\alpha|\sigma_{\dot{q}}|\beta}\braket{\beta|m}.
\end{equation}
We use Eq.(\ref{eq:tm}) to obtain the analytical results for the local temperatures shown in Fig.~\ref{fig:phasetm}c and Fig.~\ref{fig:phasetm}d.


\nocite{Danisch:2021, Rackauckas:2017, Varga:2023}

\bibliography{references}


\end{document}